\documentclass[aps,prd,superscriptaddress]{revtex4-2}

\usepackage[dvipdfmx]{graphicx}
\usepackage[dvipdfmx]{color}
\usepackage{comment}
\usepackage{amsmath,amssymb}
\usepackage{comment}
\usepackage{ulem}

\begin{document}

\title{A Search for Charged Excitation of Dark Matter with the KamLAND-Zen Detector}
\newcommand{\tohoku}{\affiliation{Research Center for Neutrino Science, Tohoku University, Sendai 980-8578, Japan}}
\newcommand{\gakusai}{\affiliation{Frontier Research Institute for Interdisciplinary Sciences, Tohoku University, Sendai 980-8578, Japan}}
\newcommand{\GPPU}{\affiliation{Graduate Program on Physics for the Universe, Tohoku University, Sendai 980-8578, Japan}}
\newcommand{\rigaku}{\affiliation{Department of Physics, Tohoku University, Sendai 980-8578, Japan}}
\newcommand{\ipmu}{\affiliation{Kavli Institute for the Physics and Mathematics of the Universe (WPI), The University of Tokyo Institutes for Advanced Study, The University of Tokyo, Kashiwa, Chiba, 277-8583, Japan}}
\newcommand{\kyoto}{\affiliation{Department of Physics, Kyoto University, Kyoto 606-8502, Japan}}
\newcommand{\osaka}{\affiliation{Graduate School of Science, Osaka University, Toyonaka, Osaka 560-0043, Japan}}
\newcommand{\rcnp}{\affiliation{Research Center for Nuclear Physics, Osaka University, Ibaraki, Osaka 567-0047, Japan}}
\newcommand{\tokushima}{\affiliation{Department of physics, Tokushima University, Tokushima 770-8506, JAPAN}}
\newcommand{\tokushimaias}{\affiliation{Graduate School of Integrated Arts and Sciences, Tokushima University, Tokushima 770-8502, Japan}}
\newcommand{\lbl}{\affiliation{Nuclear Science Division, Lawrence Berkeley National Laboratory, Berkeley, California 94720, USA}}
\newcommand{\boston}{\affiliation{Boston University, Department of Physics, 590 Commonwealth Avenue, Boston, Massachusetts 02215, USA}}
\newcommand{\hawaii}{\affiliation{Department of Physics and Astronomy, University of Hawaii at Manoa, Honolulu, Hawaii 96822, USA}}
\newcommand{\ut}{\affiliation{Department of Physics and
    Astronomy, University of Tennessee, Knoxville, Tennessee 37996, USA}}
\newcommand{\tunl}{\affiliation{Triangle Universities Nuclear
    Laboratory, Durham, North Carolina 27708, USA and \\
Physics Departments at Duke University, North Carolina Central University,
and the University of North Carolina at Chapel Hill}}
\newcommand{\virginia}{\affiliation{Center for Neutrino Physics, Virginia Polytechnic Institute and State University, Blacksburg, Virginia 24061, USA}}
\newcommand{\MIT}{\affiliation{Massachusetts Institute of Technology, Cambridge, Massachusetts 02139, USA}}
\newcommand{\nikhef}{\affiliation{Nikhef and the University of Amsterdam, Science Park, 1098XG Amsterdam, Netherlands}}
\newcommand{\washington}{\affiliation{Center for Experimental Nuclear Physics and Astrophysics, University of Washington, Seattle, Washington 98195, USA}}

% Tohoku
\author{S.~Abe}\tohoku
\author{S.~Asami}\tohoku % He joined at 2020-04-01 as M1
\author{A.~Gando}\tohoku
\author{Y.~Gando}\tohoku
\author{T.~Gima}\tohoku % She joined at 2020-04-01 as M1
\author{A.~Goto} \tohoku % She joined at 2020-04-01 as M1
\author{T.~Hachiya}\tohoku
\author{K.~Hata}\tohoku
%\author{A.~Hayashi}\tohoku
\author{S.~Hayashida}\tohoku\affiliation{Present address: Imperial College London, Department of Physics, Blackett Laboratory, London SW7 2AZ, UK}
%\author{Y.~Honda}\tohoku
\author{K.~Hosokawa}\altaffiliation[Corresponding author. ]{hosokawa@awa.tohoku.ac.jp}\tohoku
\author{K.~Ichimura} \tohoku  %joined at 2020-02-01
\author{S.~Ieki}\tohoku
\author{H.~Ikeda}\tohoku
\author{K.~Inoue}\tohoku\ipmu
\author{K.~Ishidoshiro}\tohoku
\author{Y.~Kamei}\tohoku
%\author{Y.~Karino}\tohoku
\author{N.~Kawada}\tohoku
%\author{K.~Kamizawa}\tohoku
\author{T.~Kinoshita}\tohoku
\author{M.~Koga}\tohoku\ipmu
\author{N.~Maemura}\tohoku
%\author{S.~Matsuda}\tohoku
\author{T.~Mitsui}\tohoku
\author{H.~Miyake}\tohoku
\author{K.~Nakamura}\tohoku
\author{K.~Nakamura}\tohoku % Kosuke Nakamura
\author{R.~Nakamura}\tohoku
\author{A.~Ono}\tohoku
\author{N.~Ota}\tohoku
\author{S.~Otsuka}\tohoku
\author{H.~Ozaki}\tohoku\GPPU
\author{T.~Sakai} \tohoku % joined at 2020-04-01 as M1
\author{H.~Sambonsugi}\tohoku
%\author{T.~Sato}\tohoku
%\author{Y.~Shibukawa}\tohoku
\author{I.~Shimizu}\tohoku
\author{Y.~Shirahata}\tohoku
\author{J.~Shirai}\tohoku
\author{K.~Shiraishi}\tohoku
%\author{K.~Soma}\tohoku
\author{A.~Suzuki}\tohoku
\author{Y.~Suzuki}\tohoku %Suzuki Yuya, % joined at 2020-04-01 as M1
%\author{T.~Takai}\tohoku
\author{A.~Takeuchi}\tohoku
\author{K.~Tamae}\tohoku
%\author{Y.~Teraoka}\tohoku
\author{K.~Ueshima}\tohoku\affiliation{Present address: National Institutes for Quantum and Radiological Science and Technology (QST), 1-1-1 Kouto, Sayo, Hyogo 679-5148, Japan}
\author{Y.~Wada}\tohoku
\author{H.~Watanabe}\tohoku
\author{Y.~Yoshida} \tohoku % joined at 2020-04-01 as M1

% Gakusai
\author{S.~Obara}\gakusai

% IPMU
\author{D.~Chernyak}\ipmu\affiliation{Present address: Department of Physics and Astronomy, University of Alabama, Tuscaloosa, Alabama 35487, USA and Institute for Nuclear Research of NASU, 03028 Kyiv, Ukraine}
\author{A.~Kozlov}\ipmu\affiliation{Present address: National Research Nuclear University  ``MEPhI'' (Moscow Engineering Physics Institute), Moscow, 115409, Russia}

% Osaka, RCNP
\author{S.~Yoshida}\osaka
\author{S.~Umehara}\rcnp
\author{Y.~Takemoto}\rcnp\affiliation{Present address: Kamioka Observatory, Institute for Cosmic Ray Research, The University of Tokyo, Hida, Gifu 506-1205, Japan}

% Tokushima
\author{K.~Fushimi}\tokushima
\author{S.~Hirata}\tokushimaias

%Kyoto 
\author{A.~Ichikawa}\rigaku\kyoto
\author{K.Z.~Nakamura}\kyoto
\author{M.~Yoshida}\kyoto

% LBL and UC Berkeley
%\author{T.I.~Banks}\lbl
%\author{S.J.~Freedman}\ipmu\lbl
\author{B.E.~Berger}\ipmu\lbl
\author{B.K.~Fujikawa}\ipmu\lbl
%\author{K.~Han}\lbl ?????

%Boston U
\author{C.~Grant}\boston
\author{A.~Li}\boston\affiliation{Present address: The University of North Carolina Physics \& Astronomy, 120 E. Cameron Ave., Phillips Hall CB3255, Chapel Hill, North Carolina 27599, USA}
%The University of North Carolina Physics & Astronomy, 120 E. Cameron Ave., Phillips Hall CB3255, Chapel Hill, NC 27599”
%The University of North Carolina at Chapel Hill, Chapel Hill, North Carolina 27599, USA}

%Hawaii
\author{J.G.~Learned}\hawaii
\author{J.~Maricic}\hawaii

%MIT
\author{S.~Axani}\MIT
%\author{L.~A.~Winslow}\mit
\author{L.~A.~Winslow}\MIT
\author{Z.~Fu}\MIT

% UT
\author{Y.~Efremenko}\ipmu\ut

% TUNL
\author{H.J.~Karwowski}\tunl
\author{D.M.~Markoff}\tunl
\author{W.~Tornow}\tunl

% Virginia
\author{T.~O'Donnell}\virginia
\author{S.~Dell'Oro}\virginia

% Washington
\author{J.A.~Detwiler}\washington\ipmu
\author{S.~Enomoto}\washington\ipmu

% NIKHEF
\author{M.P.~Decowski}\ipmu\nikhef

%\email[]{hosokawa@awa.tohoku.ac.jp}
%\thanks{}
%\affiliation{Research Center for Neutrino Science}

\collaboration{KamLAND-Zen Collaboration}
\noaffiliation

\date{\today}

%%%%%%%%%%%%%%%%%%%%%%%%%%%%%%%%%%%%%%%%%%%%%%%%%%%%%%%%%%%%%%%%%%%%%%%%%%%%%%%%%%%%%%%%%%%%%%%%%%%%%%%%%%%%%%%%%%%%%%%%%%%%%%%%%%%%%%%%%%%
\begin{abstract}
  There are many theories where a dark matter particle is part of a multiplet with an electrically charged state.
  If WIMP dark matter ($\chi^{0}$) is accompanied by a charged excited state ($\chi^{-}$) separated by a small mass difference, it can form a stable bound state with a nucleus.
  In supersymmetric models, the $\chi^{0}$ and the $\chi^{-}$ could be the neutralino and a charged slepton, such as the neutralino-stau degenerate model.
  The formation binding process is expected to result in an energy deposition of  {\it O}(1--10 MeV), making it suitable for detection in large liquid scintillator detectors.
  We describe new constraints on the bound state formation with a xenon nucleus using the KamLAND-Zen 400 Phase-II dataset.
  In order to enlarge the searchable parameter space, all xenon isotopes in the detector were used.
  For a benchmark parameter set of $m_{\chi^{0}} = 100$ GeV and $\Delta m = 10$ MeV, this study sets the most stringent upper limits on the recombination cross section $\langle\sigma v\rangle$ and the decay-width of $\chi^{-}$ of $2.0 \times 10^{-31}$ ${\rm cm^3/s}$ and $1.1 \times 10^{-18}$ GeV, respectively (90\% confidence level).
\end{abstract}
\maketitle

%%%%%%%%%%%%%%%%%%%%%%%%%%%%%%%%%%%%%%%%%%%%%%%%%%%%%%%%%%%%%%%%%%%%%%%%%%%%%%%%%%%%%%%%%%%%%%%%%%%%%%%%%%%%%%%%%%%%%%%%%%%%%%%%%%%%%%%%%%%
\subsection{Introduction}\label{sec:intro}
Dark matter is one of the most pressing problems in nature.
It is expected to consist of one or more new, beyond the Standard Model particles.
A well-motivated class of dark matter particles is weakly interacting massive particles (WIMPs).
The neutralino in supersymmetric (SUSY) theory is a good example of a WIMP \cite{Jungman}.
It has neutral charge and is stable relative to the age of the Universe.
Direct dark matter search experiments have searched for WIMPs interacting with nuclei with a strength weaker than the weak interaction \cite{XENON,LZ}.
Despite achieving great sensitivity to the WIMP-nucleus cross section, no signal has been observed so far.
Here, we report on an alternative search, where the WIMP is accompanied by a charged excited state.
Ref. \cite{PRL} proposed to observe the bound state formation between the charged excited state and ordinary nuclei as a method to search for this kind of dark matter particle.
The KamLAND-Zen detector has the capability to search for O(1-10) MeV energy depositions via this reaction.
In this paper, we describe new constraints on this search.

%%%%%%%%%%%%%%%%%%%%%%%%%%%%%%%%%%%%%%%%%%%%%%%%%%%%%%%%%%%%%%%%%%%%%%%%%%%%%%%%%%%%%%%%%%%%%%%%%%%%%%%%%%%%%%%%%%%%%%%%%%%%%%%%%%%%%%%%%%%
\subsection{The KamLAND-Zen 400 detector}\label{sec:KL-Zen}
The main goal of KamLAND-Zen is to search for neutrino-less double-beta decays ($0\nu\beta\beta$) in $^{136}$Xe nuclei \cite{Zen400}.
KamLAND-Zen 400 was the first phase of the project and ran from Sep. 2011 to Oct. 2015.
The KamLAND detector, located 2,700 m.w.e underground in the Kamioka mine, was originally designed to measure anti-neutrinos, such as reactor- and geo-neutrinos \cite{kamland, KLreactor, KLgeo}.
The detector is layered with an outer and an inner detector.
The outer detector (OD) is a cylindrical water Cherenkov detector to veto events primarily from cosmic muons.
The inner detector consists of a spherical stainless steel tank with 1879 Photo Multiplier Tubes (PMT) facing the inside volume, filled with organic liquid scintillator (LS) and non-scintillating mineral oil (MO).
A 6.5-m-radius spherical balloon made of nylon and EVOH inside the tank contains 1 kton of the LS (KamLS).
The remainder of the volume between the balloon and the stainless steel tank is filled with MO.
A second, 1.54-m-radius teardrop shaped balloon (inner balloon, IB) with 13 tons of xenon-loaded LS (Xe-LS) was located at the center of the KamLAND detector.
The vertex and energy are reconstructed using charge and hit timing information extracted from waveforms acquired with the PMTs.

The IB was constructed from 25 $\mu$m-thick nylon film and surrounded by the KamLS.
The Xe-LS was composed of 80.7\% decane and 19.3\% psedocumene (1,2,4-trimethylbenzene) by volume, 2.29 g/liter of the fluor PPO (2,5-diphenyloxazole) and (2.91$\pm$ 0.04)\% of xenon gas by weight.
The xenon gas was isotopically enriched and measured to be (90.77 $\pm$ 0.08)\% $^{136}$Xe and (8.96 $\pm$ 0.02)\% $^{134}$Xe.
Other xenon isotopes had negligible presence.
The energy response of the Xe-LS and the KamLS were different.
The relative light yield of the KamLS to the Xe-LS is estimated to be 1.07 \cite{sayuri}.
The difference is taken into account in this study.
The KamLAND-Zen 400 data acquisition period is divided into two phases, before and after a Xe-LS purification to handle different background rates in the Xe-LS.
Phase-I includes a background from $^{110m}$Ag ($\beta$-decay, $\tau=360$ day, $Q=3.01$ MeV) creating a peak near 2.7 MeV visible energy.
The peak-shape background disappeared after xenon distillation and LS replacement.
This analysis only uses the Phase-II dataset taken between Dec. 11$^{\rm th}$ 2013 and Oct. 27$^{\rm th}$ 2015 to avoid the peak-shaped $^{110m}$Ag background \cite{Zen400-1st}.
The livetime is 534.5 days.
The total xenon mass (all isotopes) in KamLAND-Zen 400 Phase-II was 380.7 $\pm$ 3.2 kg.

%%%%%%%%%%%%%%%%%%%%%%%%%%%%%%%%%%%%%%%%%%%%%%%%%%%%%%%%%%%%%%%%%%%%%%%%%%%%%%%%%%%%%%%%%%%%%%%%%%%%%%%%%%%%%%%%%%%%%%%%%%%%%%%%%%%%%%%%%%%
\subsection{The model and the expected signal}
\begin{figure}[t]
  \begin{center}
    \includegraphics[width=22pc]{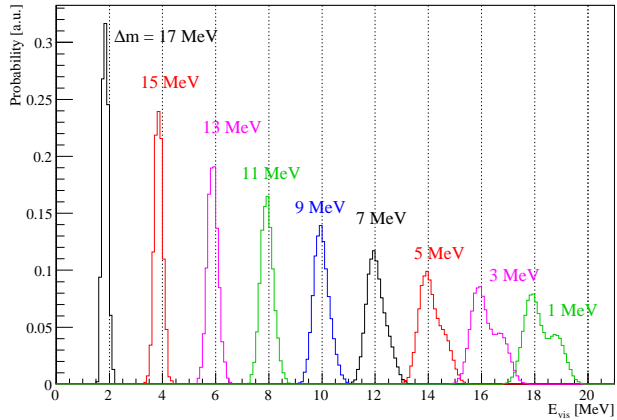}
    \caption{\label{signal}
      The expected energy spectra via the bound state formation for several $\Delta m$ in the KamLAND-Zen 400 detector.}
  \end{center}
\end{figure}

As mentioned in Section \ref{sec:intro}, there are scenarios where a WIMP is part of a multiplet with an electrically charged excited state \cite{PRL, Maxim2008}.
If such a WIMP does not have a large enough annihilation cross-section, co-annihilation only can determine the dark matter's abundance.
If the mass difference $\Delta m$ between the WIMP and the excited state is sufficiently small, the charged excitation of the WIMP can form a stable bound state with a nucleus \cite{PRL}.
In this process, the Coulomb binding energy, whose value depends on the target nucleus, is released and the observable energy is of the order {\it O}(1--10 MeV).
Detectors studying neutrinos and $0\nu\beta\beta$ are well-suited to detect events in this energy range.
The bound state formation process and the effective Yukawa-type Lagrangian, which governs the bound state formation are given by \cite{PRL}(case A in the reference):
\begin{align}
  N_Z + \chi^0 \rightarrow (N_Z\chi^-) + e^+,\\
  \mathcal{L} = \bar{\chi^{0}}(g_{eL}{\bf P}_{L} + g_{eR}{\bf P}_{R})e {\chi^{-}}^{\dagger} + {\rm H.c.},
\end{align}
where $N_Z$ is the target nucleus with atomic number {\it Z}, $g_{eL,R}$ are the general complex couplings of the chirality projections ${\bf P}_{L,R}$, and $\chi^0$ ($\chi^-$) is the WIMP ground (excited) state, respectively.
In SUSY models, the latter two could be the neutralino and a charged slepton, such as the neutralino-stau degenerate model \cite{Ellis1998,Ellis2000}.
If the bound state $(N_Z\chi^-)$ is not in its ground state, it will de-excite by emitting $\gamma$-rays.
  In addition to the de-excitation $\gamma$-rays and a positron, annihilation $\gamma$-rays could be observed in this process.
  Therefore, the observable energy is given by \cite{PRL}:
\begin{align}
  E_{\rm tot} &= E_{e^+} + E_{\gamma} + 2m_{e},
\end{align}
where
\begin{align}
  E_{e^+} &= E_{b}^{(n,l)} - \Delta m - m_{e},\\
  E_{\gamma} &= E_{b}^{(0)} - E_{b}^{(n,l)},
\end{align}
so that
\begin{align}
  E_{\rm tot} &= E_{b}^{(0)} - \Delta m + m_{e},
\end{align}
where $E_{e^+}$ and $E_{\gamma}$ are the kinetic energy of the positron and the de-excitation $\gamma$-rays.
The Coulomb binding energy $E_{b}$ of $(N_Z\chi^-)$ allows to bridge the mass difference $\Delta m \equiv m_{\chi^-} - m_{\chi^0}$, the value depends on the target nucleus.
Assuming a step-like nuclear charge distribution, $E_{b}^{(0)}$, the binding energy corresponding to the ground state of $(N_Z\chi^-)$, is calculated to be 18.4 MeV for a xenon target \cite{Maxim2008}.
By increasing Z, $E_{b}^{(0)}$ also increases and the searchable $\Delta m$ region is larger, so that xenon nuclei offer a good target.
$E_{b}^{(n,l)}$ is the excited-state energy with the usual principal and orbital quantum numbers of the capture level ({\it n,l});
The energy distributions of the positrons and the $\gamma$-rays change with $E_{b}^{(n,l)}$.
However, $E_{\rm tot}$ is monochromatic, regardless of the capture level.
The signal shape is basically determined only by the energy response of the detector.
Figure \ref{signal} shows the expected energy spectra for several $\Delta m$ values, where $E_{\rm vis}$ is the visible energy in KamLAND-Zen and differs from the intrinsic energy deposition due to energy non-linearity and the light yield difference between the Xe-LS and the KamLS.
The energy non-linearity, the light yield difference and the energy resolution ($\sigma_E = \rm 7.3\% / \sqrt{{\it E} (MeV)}$) are taken into account \cite{Zen400}.
The higher and lower visible energy ranges correspond to lower and higher $\Delta m$, respectively.
Here a single nuclear de-excitation $\gamma$-ray with total energy $E_{\gamma}$ is considered; multiple $\gamma$-rays emissions give a negligible difference from the single gamma-ray analysis.

The expected number of signals in an energy bin $\Delta E = E_{\rm max} - E_{\rm min}$ is given by \cite{PRL}:
\begin{align}
  \label{eq:Nsignal}
  N_{\rm expected} = \frac{MTN_{T}\rho_{\rm DM} \langle\sigma v\rangle}{2m_{\chi^0}}
  \left[
    {\rm Erf}\left(\frac{E_{\rm max}-E_{\rm tot}}{\sqrt{2}\sigma_{E}}\right) - {\rm Erf}\left(\frac{E_{\rm min}-E_{\rm tot}}{\sqrt{2}\sigma_{E}}\right)
    \right],
\end{align}
where {\it MT}, $N_{T}$, $\rho_{\rm DM}$ and $\langle\sigma v\rangle$ are detector exposure, number of target nuclei, the local density of dark matter and the WIMP-nucleus recombination cross section with incoming dark matter velocity {\it v}, respectively.
In general, the signal is expected to have a Gaussian peak.
However, the simulated signal shapes in the KamLAND-Zen 400 detector (Fig.\ref{signal}) have non-Gaussian distributions due to the light yield difference between the Xe-LS and the KamLS.
The $\Delta E$ has therefore to be defined considering the non-Gaussian signal shape.
Once the WIMP mass $m_{\chi^0}$ and $\Delta m$ are chosen, the induced detector signal can be translated into a constraint on $\langle\sigma v\rangle$ or the $\chi^{-}$'s decay width $\Gamma_{\chi^{-}}$, following Eq.(\ref{eq:Nsignal}) \cite{PRL}:
\begin{align}
  \langle\sigma v\rangle &\simeq (|g_{eL}|^2 + |g_{eR}|^2) / (8\pi m_{\chi^{0}}) \times \sum_{n,l} B_{n,l},\\
  \label{eq:Bnl}
  B_{n,l} &\simeq (E_b^{(n,l)} - \Delta m - m_e) \sqrt{( E_b^{(n,l)} - \Delta m )^2 - m_e^2} \times
  \int d^3r_1 d^3r_2 \phi^{\ast}_{n,l}(\vec{r}_1)\phi_{n,l}(\vec{r}_2)e^{i\mu\vec{v}\cdot(\vec{r}_1-\vec{r}_2)},\\
  \Gamma_{\chi^{-}} &= \tau_{\chi^{-}}^{-1}\\
  &\simeq \frac{\sqrt{\Delta m^{2} - m_{e}^{2}}}{4\pi m_{\chi^{0}}} (\Delta m + m_{e}) (|g_{eL}|^2 + |g_{eR}|^2),
\end{align}
where {\it $B_{n,l}$}, $\phi_{n,l}$ and $\tau_{\chi^{-}}$ are the contributions from the capture into the state ({\it n,l}), the wave function of the relative motion of ($N\chi^{-}$) with reduced mass $\mu$ and the lifetime of $\chi^{-}$, respectively.
A translation into a constraint on $\Gamma_{\chi^{-}}$ allows us to combine our results with the limits on the stau's decay width $\Gamma_{\tilde\tau}$ obtained in collider experiments, such as CMS \cite{CMS}.

%%%%%%%%%%%%%%%%%%%%%%%%%%%%%%%%%%%%%%%%%%%%%%%%%%%%%%%%%%%%%%%%%%%%%%%%%%%%%%%%%%%%%%%%%%%%%%%%%%%%%%%%%%%%%%%%%%%%%%%%%%%%%%%%%%%%%%%%%%%
\subsection{Analysis}\label{sec:anal}
As a preliminary step to the analysis, we apply the same event selection to the dataset as for the $0\nu\beta\beta$ analysis \cite{Zen400}:
\begin{enumerate}
\item
  Events with energies over $\sim 30$  MeV and OD-triggered events are vetoed as muon events.
  Events within 2 ms after a muon crossing are also vetoed to reject PMT afterpulses and instability of signal baseline;
\item
  Sequential decays of $^{214}$Bi--Po and $^{212}$Bi--Po are tagged by {\it delayed-coincidence} and {\it pileup-detection} methods and rejected;
\item
  Events from shorter-lived spallation products {\it i.e.} $^6$He, $^8$Li, $^{10}$C and $^{12}$B, are suppressed by a triple-coincidence tag of a muon, a neutron identified by capture $\gamma$ rays and the product's decay;
\item
  Anti-neutrino events which produce a positron and a neutron from inverse $\beta$-decay are identified with the delayed-coincidence tagging method and vetoed;
\item
  Badly reconstructed events are identified by time-charge-hit discriminator and rejected.
\end{enumerate}
The total detection efficiency for the bound state formation is over 99.9\%, where the deadtime due to the spallation cut and post-muon events are taken into account in the livetime calculation.

\begin{figure}[t]
  \begin{center}
    \includegraphics[width=22pc]{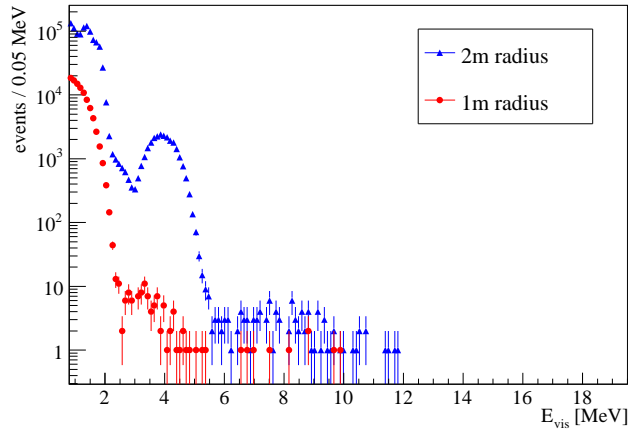}
    \caption{\label{observed}
      The observed energy spectra in the KamLAND-Zen 400 Phase-II dataset for a 1 m and 2-m-radius fiducial volume.
    }
  \end{center}
\end{figure}

Figure \ref{observed} shows the observed energy spectrum, which includes the high-energy range (above 4.8 MeV) not used in the $0\nu\beta\beta$ analysis for two different fiducial volumes.
The two-neutrino double-beta decay ($2\nu\beta\beta$) events dominate the energy region.
The Q-value of $^{136}$Xe $2\nu\beta\beta$ is 2.458 MeV and the spectrum has a continuous distribution.
The IB is not as clean as the Xe-LS and contains radioactive sources from the $^{232}$Th and $^{238}$U decay series.
Therefore, the background rate with a 2-m-radius fiducial volume is much higher than with a 1-m-radius.
The peak at 4 MeV in the 2-m-radius analysis is dominated by $\beta + \gamma$-rays (Q-value $\sim$ 5 MeV) from $^{208}$Tl decay contained in the IB.
The high energy region above 5 MeV is dominated by short-lived spallation events from cosmic muons such as $^{12}$B, $^{8}$Li and $^{8}$B.
The high energy detector response is understood within 1\% uncertainty in the KamLS by using spallation events just after muon events \cite{sayuri, obara}.

In order to optimize the fiducial volume, the radius was selected using a figure of merit (FoM), which is a function of the energy range:
\begin{align}
  {\rm FoM}(r,\Delta m) \equiv \frac{S}{\sqrt{B}} \equiv \frac{{\rm FV}(r) \times \epsilon _{\rm det} (r,{\rm \Delta m})}{\sqrt{\rm {\it N}{_{\ obs}^{\ 90\%}}}},
\end{align}
where FV({\it r}) is the volume of the Xe-LS, ${\rm {\it N}_{\ obs}^{\ 90\%}}$ is the 90\% confidence level (C.L.) upper limit on the number of observed events.
The reconstructed position of the expected signal is spatially diffused due to the emitted $\gamma$-rays.
The spatial detection efficiency $\epsilon _{\rm det} (r,{\rm \Delta m})$ is estimated from a Monte-Carlo simulation.
Similarly to the $0\nu\beta\beta$ analysis, this analysis uses the volume within 2-m-radius from the center of the IB and includes both the internal and external regions of the Xe-LS and the IB.
The volume was divided into 20 equal-volume spherical shell bins.
The FoM was then calculated for each of the 20 equal-volume bins and the bin with the highest FoM was chosen at 1 MeV $\Delta m$ intervals.

%%%%%%%%%%%%%%%%%%%%%%%%%%%%%%%%%%%%%%%%%%%%%%%%%%%%%%%%%%%%%%%%%%%
% ROI energy width optimization
As shown in Figure \ref{signal}, the width of the expected spectrum changes with the value of $\Delta m$.
To improve the S/N ratio, the width of the energy region of interest was defined to be twice the standard deviation of the expected signal.
This region includes over 95\% of the expected signal.

%%%%%%%%%%%%%%%%%%%%%%%%%%%%%%%%%%%%%%%%%%%%%%%%%%%%%%%%%%%%%%%%%%%
% Known background subtraction
To improve the sensitivity, the 90\% upper limit of the signal rate was calculated from the number of observed and known-background events in the energy range of interest using the Feldman and Cousins method \cite{FeldmanAndCousins}.
The best-fit $2\nu\beta\beta$ spectrum and the well-known background spectra fixed in the $0\nu\beta\beta$ analysis were used.
The fixed backgrounds are ($^{232}$Th series, $^{210}$Bi, $^{85}$Kr) in the Xe-LS, ($^{232}$Th series, $^{210}$Bi, $^{85}$Kr, $^{40}$K) in the KamLS and ($^{232}$Th series, $^{210}$Bi, $^{137}$Cs, $^{134}$Cs, $^{40}$K) contained on the IB film.

%%%%%%%%%%%%%%%%%%%%%%%%%%%%%%%%%%%%%%%%%%%%%%%%%%%%%%%%%%%%%%%%%%%
\subsection{Result and conclusion}
% Result
\begin{figure}[t]
  \begin{center}
    \includegraphics[width=22pc]{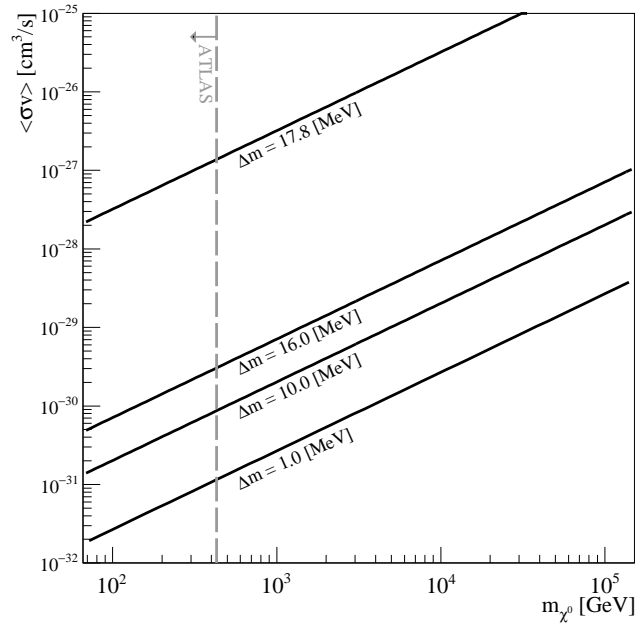}
    \caption{\label{resultsigmav}
      The $N_z$-$\chi^{0}$ recombination cross section $\langle\sigma v\rangle$ as a function of the WIMP mass $m_{\chi^{0}}$.
      The black solid lines show the 90\% C.L. upper limit from this study for several $\Delta m$ values.
      The gray dashed line shows the lower limit for the mass of the $\tilde\tau$ from the ATLAS experiment \cite{ATLAS}.
    }
  \end{center}
\end{figure}
\begin{figure}[t]
  \begin{center}
    \includegraphics[width=22pc]{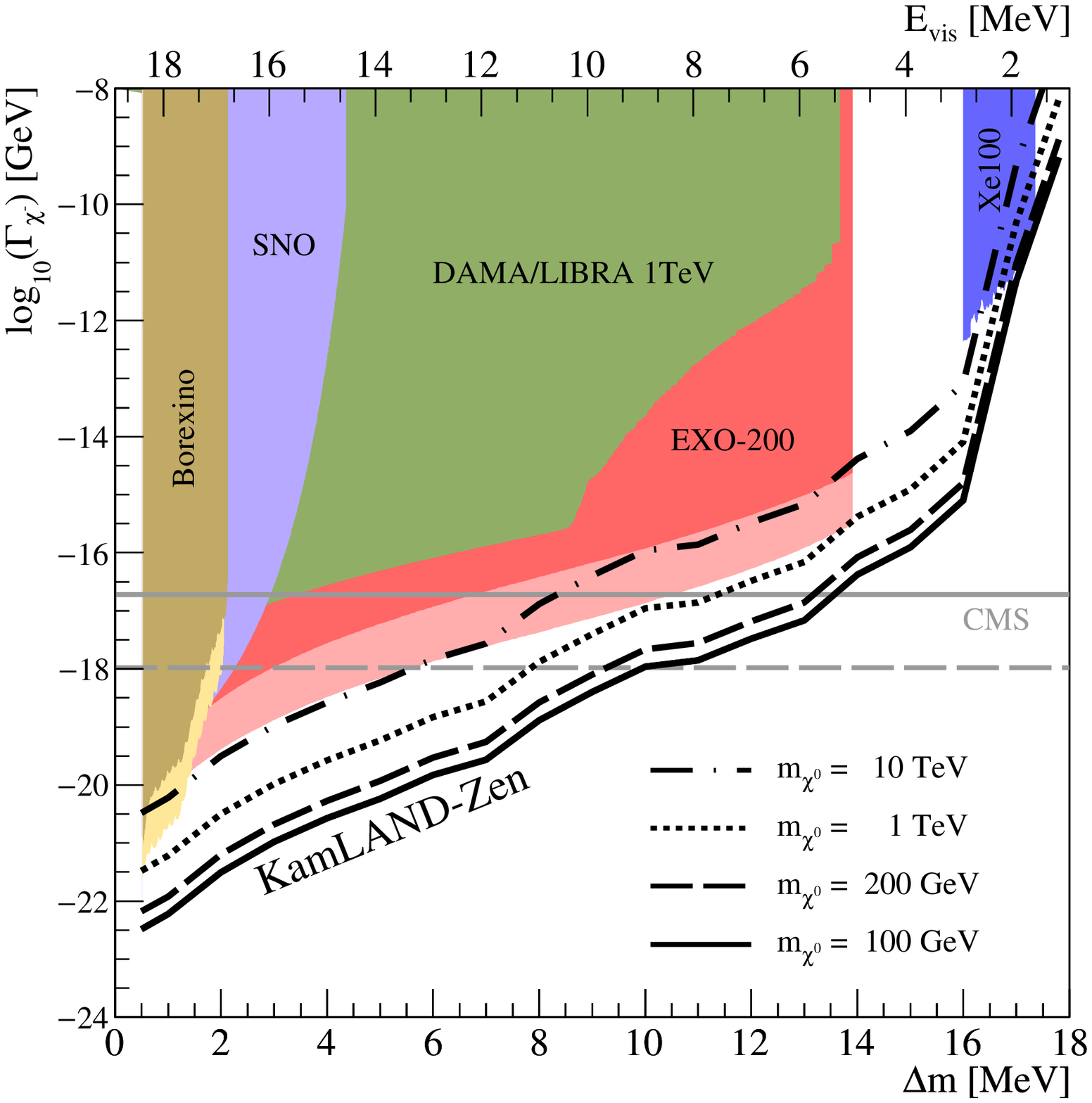}
    \caption{\label{result}
      Excluded decay width region of $\tilde\tau$ as a function of $\Delta m$ (bottom axis) and $E_{vis}$ (top axis).
      The black curves show 90\% C.L. upper limits from the KamLAND-Zen 400 Phase-II dataset.
      Horizontal solid (dashed) lines show the lower limit assuming a $\tilde\tau$ mass of 100 (200) GeV by the CMS experiment \cite{PRL,CMS}.
      Filled regions are theoretical constraints using reported spectra from several experiments \cite{PRL}.
      The darker and lighter regions correspond to constraints for 1 TeV and 100 GeV, respectively.
    }
  \end{center}
\end{figure}

The results of this study for the WIMP parameter space are shown in Figure \ref{resultsigmav}.
No significant excess of events was found.
The upper limit for the $N_{Z}$-$\chi^{0}$ recombination cross section $\langle\sigma v\rangle$ is drawn as a function of $m_{\chi^0}$ for several $\Delta m$ values, including a lower limit for the $\tilde\tau$ mass obtained by the ATLAS experiment \cite{ATLAS}.
Figure \ref{result} shows another interpretation of the results.
The black curves are the upper limits on the $\Gamma_{\chi^{-}}$ by KamLAND-Zen 400 Phase-II.
The curves corresponds to $m_{\chi^0}$ of 100 GeV, 200 GeV, 1 TeV and 10 TeV respectively.
Theoretical constraints based on reported spectra of several experiments \cite{PRL} are also shown.
As are lower limits from a search for $\tilde\tau$ by the CMS assuming the minimal gauge-mediated super-symmetry breaking (GSMB) model \cite{PRL,CMS}.

%%%%%%%%%%%%%%%%%%%%%%%%%%%%%%%%%%%%%%%%%%%%%%%%%%%%%%%%%%%%%%%%%%%%%%%%%%%%%%%%%%%%%%%%%%%%%%%%%%%%%%%%%%%%%%%%%%%%%%%%%%%%%%%%%%%%%%%%%%%
%\subsection{Conclusion}
A search for the bound state formation of a nucleus and an electrically charged WIMP state was performed using data from the $0\nu\beta\beta$ detector KamLAND-Zen 400.
The analysis has good sensitivity in the region $\Delta m < 17.8 {\rm\ MeV}$ and access to higher slepton masses than accessible at the Large Hadron Collider \cite{ATLAS} when a SUSY scenario is assumed.
The upper limit for the recombination cross section is $\langle\sigma v\rangle < 2.0 \times 10^{-31}$ ${\rm cm^3/s}$ when assuming $\Delta m = 10 {\rm\ MeV}$ and $m_{\chi^{0}} = 100$ GeV.
The upper limit of the decay width of $\chi^{-}$ assuming $\Delta m = 10 {\rm\ MeV}$ is $\Gamma_{\chi^{-}} < 1.1 \times 10^{-18} {\rm\ GeV}$ and $\Gamma_{\chi^{-}} < 1.1 \times 10^{-16} {\rm\ GeV}$ for $m_{\chi^{0}} = 100 {\rm\ GeV}$ and $m_{\chi^{0}} = 10 {\rm\ TeV}$, respectively.

The KamLAND-Zen 800 experiment, the next phase of KamLAND-Zen 400, has recently started acquiring data with about twice as much xenon and a new IB with ten times lower radioactive contamination \cite{TAUP2019}.
It is expected to improve the sensitivity and decrease the $^{208}$Tl background at 4 MeV originating from the IB.
However, spallation events from cosmic muons in the high-energy region over 5 MeV will become a challenging background.

%%%%%%%%%%%%%%%%%%%%%%%%%%%%%%%%%%%%%%%%%%%%%%%%%%%%%%%%%%%%%%%%%%%%%%%%%%%%%%%%%%%%%%%%%%%%%%%%%%%%%%%%%%%%%%%%%%%%%%%%%%%%%%%%%%%%%%%%%%%
\begin{acknowledgments}
  The authors wish to acknowledge Dr. Haipeng An and Dr. Josef Pradler for providing helpful advice and the values related to the contribution from a capture into a state ({\it n,l}) in equation \ref{eq:Bnl}.
  The KamLAND-Zen experiment is supported by JSPS KAKENHI Grants
  No. 21000001,
  No. 26104002,
  and No. 19H05803;
  the World Premier International Research Center Initiative (WPI Initiative), MEXT, Japan;
  Netherlands Organization for Scientific Research (NWO) ;
  and under the U.S. Department of Energy (DOE) Contract No. DE-AC02-05CH11231,
  the National Science Foundation (NSF) No. NSF-1806440,
  as well as other DOE and NSF grants to individual institutions.
  The Kamioka Mining and Smelting Company has provided service for activities in the mine.
  We acknowledge the support of NII for SINET4. 
\end{acknowledgments}

\bibliography{hoge.bib}
\bibliographystyle{unsrt}

\end{document}